# Meme creation and sharing processes: individuals shaping the masses

IAN MILLER and GERALD CUPCHIK, University of Toronto

## 1. INTRODUCTION

Individuals are able to effect massive online social change through the viral sharing of memes. Sharing is initially influenced by meme creators and secondarily by meme consumers, whose individual sharing decisions accumulate to determine total meme sharing. By investigating meme creation as a separate process from meme sharing, we hope to create an account of online sharing that includes features of both processes. The present work uses psychological methods, web log monitoring [Harley et al. 2006], statistical modeling, and agent modeling [North et al. 2006] to explore meme creation and sharing. Over the course of two studies, human participants were observed creating and sharing memes during a 70-day period, and these behaviors were statistically modeled. The resulting model of sharing was implemented as an agent simulation to investigate whether models of contagion [Dodds and Watts 2005; Christakis and Fowler 2012] could simulate the dynamics of our human sample.

## 2. CONTENT CREATION AND SHARING (STUDY 1)

There are two phases to this study: (1) observing the meme creation process; and (2) monitoring the online sharing of those memes during a 70-day period. Our investigation is restricted to a specific set of memes known as *image macros*, which are digital images that have captions embedded along the top and bottom of the image. Image macros may also be called LOLcats, advice animals, or even simply *memes*. Unlike text-mediated memes (e.g. Twitter) which are subject to rapid mutation [Leskovec et al. 2009; Kooti et al. 2012], images macros are more difficult to alter during the sharing process. Commercial image macro builder websites (e.g. quickmeme.com) make it simpler to generate a new image macro than to modify an existing one. Each time an image macro is generated, a unique URL is typically assigned to the image, which may then be used to share the image macro online. Due to the strong mapping between the URL and a specific image macro, this type of meme is relatively easy to monitor.

### 2.1 Methods

118 participants from an undergraduate research pool at a large, North American university participated for course credit. The sample was predominantly female (76.52%). Approximately half of our participants had previously created and shared memes before coming into the lab.

An online meme creating and sharing website (internally called *memelab*) was built to support this investigation. The corresponding website is a publicly accessible, memorable 5-letter Dot-com providing a user experience that closely mimics other popular meme sharing websites. The web application, web server, and access logs were hosted on a university computer located in a secure lab room.

In order to observe the creation process, we invited participants to our campus laboratory. The laboratory space consists of several individual student workstations separated by cubicle partitions. The workstation was prepared for participants by loading the *memelab* website ahead of time.

2.1.1 *Meme rating task.* At the beginning of the laboratory session, participants completed a series of surveys to capture demographic information and Internet usage profiles. Other surveys recorded participant responses to a variety of psychological measures that would be used during modeling.

Next, participants completed the meme rating task. In order to build a model of meme sharing decisions, we asked participants to look at 8 memes and rate them. The ratings included a measure of how likely they would be to share the meme online. Meme relevance was manipulated to create high-relevance and low-relevance conditions by showing 4 academic themed image macros and 4 animal themed image macros, respectively.

2.1.2 *Creation Process.* As the last part of the lab session, participants engaged in an image macro creation task. A 30-second demonstration video provided an introduction to *memelab* website operation. During this task, participants picked a background picture from our library of common meme backgrounds and added captions to the top and bottom of the picture. The captions were rendered directly into the image, thereby creating a unique, immutable meme that could be shared online. The meme was assigned a short, permanent, unique URL. At the end of the study, participants received a hard copy record of their meme URLs for future reference. Each participant created two memes for a total sample of 236 memes.

2.1.3 *Sharing Process.* The "memelab" creation process naturally flows into the sharing process. The last screen of the meme creation task is called the "sharing screen". In addition to displaying the image macro itself, the sharing screen includes linking information and a brief guide to embedding the URL in social media messages.

Any time a new user viewed a meme, they too would see the sharing screen, enabling a "snowball effect". We monitored the sharing process from our web server, which created a log entry when one of the study's memes was requested by any web browser. These access requests were used as a proxy for sharing behavior. Typically, the only times those URLs were accessed was because the meme had been shared with a new user.

## 2.2 Results

We monitored traffic over a 70-day period, during which we logged approximately 2000 total requests for the study's 236 memes (see figure 1, left panel). Most memes were viewed fewer than twice, while the top-viewed meme was requested 72 times. Based on psychometric and sharing data, we used a guided model fitting procedure to derive two statistical models:

2.2.1 *Creator model of re-sharing behavior.* The creator model of re-sharing behavior predicts total re-sharing behavior (i.e. total hits received by a meme) based on meme ratings including humor, self-relevance, predicted-liking, and multivariate psychological scales. Although this model only contained information about the creator and the meme, it explained 11.5% of the variance in consumer sharing behavior.

2.2.2 *Consumer model of meme sharing decisions.* The consumer model of individual sharing decisions predicts the individual decision to share a meme based on features of the meme: (1) humor ratings; (2) self-relevance ratings; and (3) self-reference words in captions. The resulting model explains 37.5% of the variance in sharing decisions.

## 3. A CONTAGION MODEL OF MEME SHARING (STUDY 2)

The next stage of this investigation sought to replicate human data using our model of meme sharing decisions. We implemented our sharing model as the infection function in an SIS simulation of contagion (i.e. agents are either susceptible or infected). In choosing parameters for the model, we sought to match the conditions of the human study as closely as possible. As there were 15,000 students on the campus where the study was conducted, there were also 15,000 agents in the simulation. Since we recruited 118 students, the simulation also recruited 118 agents and created 238 total memes. We ran the study an average of once every 4 days, so the simulation waited 4 time ticks between recruiting agents.

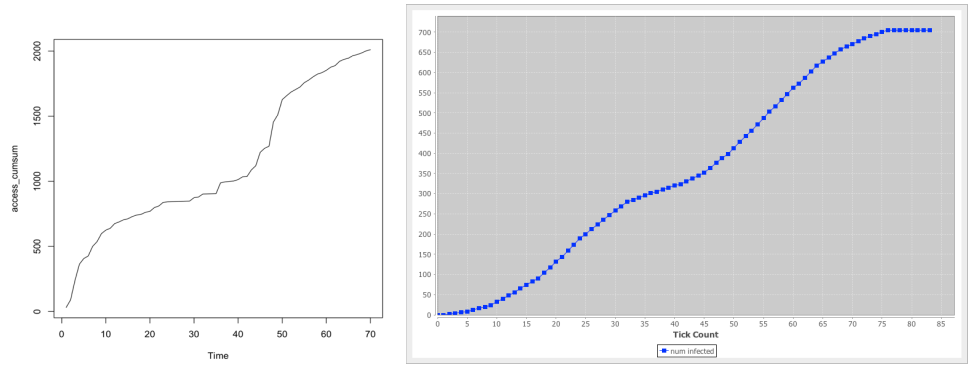

Fig. 1. These time series compare human sharing results (left) and simulation results (right).

3.1 Methods

The Repast Symphony IDE [North et al. 2013] was used to construct an environment in which to simulate our campus population. Agents were randomly assigned locations within a 2-dimensional space. A random walk through this space caused agents to unpredictably encounter one another, which is analogous to the unpredictable nature of encountering other users in large online forums.

The experimenter was represented as a zombie character who periodically recruited agents to participate in the study. Our simulation represented meme content as a vector of values chosen randomly from the Gaussian distribution. An agent received a random meme vector when they were recruited by the experimenter. When an agent possessed a meme, they used our sharing model to decide whether to share with nearby agents. If an agent decided to share, nearby agents became infected for a certain period of time. Due to the random walk dynamic, other agents were infected and the memes propagated. This simulation tracked the total number of infections over time.

3.2 Results

In the framework of the contagion model, total infections over time is analogous to total re-sharing among human participants. A qualitative comparison of web traffic data with our simulation time series (see figure 1, right panel yields several encouraging similarities, such as an adoption curve. Future research experimentally investigate whether simulation results can be replicated in a human sample.